\documentclass{llncs}
\usepackage{amsmath}
\usepackage{amssymb}
\usepackage{wrapfig}

\usepackage{cutwin}

\PassOptionsToPackage{hyphens}{url}\usepackage{hyperref}
\usepackage{graphicx}
\usepackage[ruled,vlined]{algorithm2e}
\usepackage{bm}
\usepackage{xspace}
\usepackage{subfig}

\DeclareMathOperator{\Inf}{Inf}

\DeclareMathAlphabet{\mathbfsf}{\encodingdefault}{\sfdefault}{bx}{n}

\makeatletter
\def\wrapfill{\par
   \ifx\parshape\WF@fudgeparshape
     \nobreak
     \ifnum\c@WF@wrappedlines>\@ne
       \advance\c@WF@wrappedlines\m@ne
       \vskip\c@WF@wrappedlines\baselineskip
       \global\c@WF@wrappedlines\z@
     \fi
     \allowbreak
     \WF@finale
   \fi
}
\makeatother

\begin{document}

\newcommand{\hgrant}{\textit{hgrant}}
\newcommand{\hbusreq}{\textit{hbusreq}}
\newcommand{\alarm}{\textit{alarm}}
\newcommand{\stopp}{\textit{stop}}
\newcommand{\DP}{\textit{DP}}
\newcommand{\stateone}{\textit{state1}}
\newcommand{\statetwo}{\textit{state2}}
\newcommand{\statethree}{\textit{state2}}
\newcommand{\EAT}{\textit{EAT}}
\newcommand{\kh}{\textit{kh}}
\newcommand{\Inff}{\textit{Inf}}

\newcommand{\stepi}{\textit{i}}
\newcommand{\ii}{\textit{ii}}
\newcommand{\iii}{\textit{iii}}
\newcommand{\iv}{\textit{iv}}
\newcommand{\old}{\textit{old}}

\renewcommand\topfraction{0.85}
\renewcommand\bottomfraction{0.85}
\renewcommand\textfraction{0.20}
\renewcommand\floatpagefraction{0.85}
\setlength{\intextsep}{1\baselineskip}
\newcommand{\hmaster}{\textit{hmaster}}
\newcommand{\hready}{\textit{hready}}
\newcommand{\hbusreqone}{\textit{hbusreq1}}
\newcommand{\hburstzero}{\textit{hburst0}}
\newcommand{\hburstone}{\textit{hburst1}}
\newcommand{\hmasterzero}{\textit{hmaster0}}
\newcommand{\hmastlock}{\textit{hmastlock}}
\newcommand{\hbusreqzero}{\textit{hbusreq0}}
\newcommand{\hgrantzero}{\textit{hgrant0}}
\newcommand{\hgrantone}{\textit{hgrant1}}
\newcommand{\hlockzero}{\textit{hlock0}}
\newcommand{\hlockone}{\textit{hlock1}}
\newcommand{\stateG}{\textit{stateG}}
\newcommand{\stateA}{\textit{stateA}}
\newcommand{\start}{\textit{start}}
\newcommand{\decide}{\textit{decide}}
\newcommand{\locked}{\texts{locked}}
\newcommand{\inx}{\mathcal{X}}
\newcommand{\minx}{$\mathcal{X}$\xspace}
\newcommand{\outy}{\mathcal{Y}}
\newcommand{\mouty}{$\mathcal{Y}$\xspace}
\newcommand{\var}{\mathcal{V}}
\newcommand{\mvar}{$\mathcal{V}$}

\newcommand{\lang}{\mathcal{L}}
\newcommand{\tran}{\mathcal{T}}
\newcommand{\interpolant}{\mathcal{I}}
\newcommand{\game}{\mathcal{G}}
\newcommand{\mgame}{$\mathcal{G}$}

\newcommand{\msys}{$\mathcal{S}$}
\newcommand{\sys}{\mathcal{S}}
\newcommand{\mcont}{$\mathcal{C}$}
\newcommand{\cont}{\mathcal{C}}
\newcommand{\menv}{$\mathcal{E}$}
\newcommand{\env}{\mathcal{E}}
\newcommand{\win}{\textit{Win}}

\newcommand{\fair}{{\scriptsize\textit{fair}}}
\newcommand{\inv}{{\scriptsize\textit{inv}}}
\newcommand{\init}{{\scriptsize\textit{init}}}

\newcommand{\trans}{\textit{trans}}
\newcommand{\loopp}{\textit{loop}}
\newcommand{\fail}{\textit{fail}}
\newcommand{\unr}{\textit{unr}}
\newcommand{\cfair}{\textit{cfair}}

\newcommand{\true}{\textit{true}}
\newcommand{\false}{\textit{false}}

\newcommand{\always}{\mathbfsf{G}}
\newcommand{\eventually}{\mathbfsf{F}}
\newcommand{\next}{\mathbfsf{X}} 
\newcommand{\until}{\mathbfsf{U}}

\newcommand{\initstates}{S^{\init}}
\newcommand{\transstates}{S^{\trans}}
\newcommand{\loopingstates}{S^{\loopp}}
\newcommand{\failingstates}{S^{\fail}}
\newcommand{\unrolledstates}{S^{\unr}}
\newcommand{\statesu}{S_u}

\newcommand{\initstate}{s^{\init}}
\newcommand{\failingstate}{s^{\fail}}
\newcommand{\transstate}[1]{s^{\trans}_{#1}}
\newcommand{\loopingstate}[1]{s^{\loopp}_{#1}}
\newcommand{\unrolledstate}[2]{s^{\unr}_{#1,#2}}

\newcommand{\tcounterplay}{[[\pi_{C,u},\varphi^\env_u]]}
\newcommand{\tcassumptions}{[[\varphi^\env_u]]}
\newcommand{\tcguarantees}{[[\varphi^\sys_u]]}

\newcommand{\boolexpr}[3]{B^{\text{#1}}_{#2}(#3)}
\newcommand{\boolxproj}[3]{B^{\text{#1}}_{\ifthenelse{\equal{#2}{}}{\inx}{#2,\inx}}(#3)}
\newcommand{\boolyproj}[3]{B^{\text{#1}}_{\ifthenelse{\equal{#2}{}}{\outy}{#2,\outy}}(#3)}
\newcommand{\boolexprvar}{\boolexpr{}{}{\var}}
\newcommand{\boolexpression}{B}

\newcommand{\assumptionsinitset}{\Psi_{\init}}
\newcommand{\assumptionsinvset}{\Psi_{\inv}}
\newcommand{\assumptionsfairset}{\Psi_{\fair}}

\newcommand{\olanguage}{$\omega$-\-lan\-guage\xspace}
\newcommand{\olanguages}{$\omega$-\-lan\-gua\-ges\xspace}
\newcommand{\oword}{$\omega$-\-word\xspace}
\newcommand{\owords}{$\omega$-\-words\xspace}
\newcommand{\sigmaomega}{\Sigma^\omega}
\newcommand{\sigmastar}{\Sigma^*}
\newcommand{\sigmavar}{\Sigma_\var}

\newcommand{\bautomaton}{\mathcal{B}}
\newcommand{\mautomaton}{\mathcal{M}}
\newcommand{\lautomaton}[1]{\mathcal{A}_{#1}}

\newcommand{\Hausdim}[1]{\dim\left({#1}\right)}
\newcommand{\Hausmeas}[2]{m_{#1}({#2})}
\newcommand{\HausmeasBf}[2]{\mathbf{Haus}_{#1}({#2})}

\newcommand{\preflanguage}[1]{{A({#1})}}
\newcommand{\prefnlanguage}[2]{{A_{#1}({#2})}}
\newcommand{\suff}[2]{\operatorname{S}_{#1}({#2})}
\newcommand{\suffixlanguage}[2]{{\suff{#1}{#2}}}

\newcommand{\entropy}[1]{H({#1})}

\newcommand{\closure}{\mathcal{C}}
\newcommand{\setsize}[1]{{\#({#1})}}

\newcommand{\initial}{\phi^{\init}}
\newcommand{\invariant}{\phi^{\inv}}
\newcommand{\fairness}{\phi^{\fair}}
\newcommand{\cfairness}{\phi^{\cfair}} 

\newtheorem{axiom}{Axiom}

\mainmatter              
\title{A Weakness Measure for GR(1) Formulae}
\titlerunning{A Weakness Measure for GR(1) Formulae}  
%
\author{Davide G. Cavezza \and Dalal Alrajeh \and Andr\'as Gy\"orgy}
\authorrunning{D.~Cavezza, D.~Alrajeh and A.~Gy\"orgy} 
\institute{Imperial College London, London, United Kingdom\\
\email{\{d.cavezza15,dalal.alrajeh,a.gyorgy\}@imperial.ac.uk}
}

\maketitle             




\begin{abstract}
In spite of the theoretical and algorithmic developments for system synthesis in recent years, little effort has been dedicated to quantifying the quality of the specifications used for synthesis.
When dealing with unrealizable specifications, finding the weakest environment assumptions that would ensure realizability is typically a desirable property; in such context the weakness of the assumptions is a major quality parameter. The question of whether one assumption is weaker than another is commonly interpreted using implication or, equivalently, language inclusion. However, this interpretation does not provide any further insight into the weakness of assumptions when implication does not hold.
To our knowledge, the only measure that is capable of comparing two formulae in this case is entropy, but even it fails to provide a sufficiently refined notion of weakness in case of GR(1) formulae, a subset of linear temporal logic formulae which is of particular interest in controller synthesis. In this paper we propose a more refined measure of weakness based on the Hausdorff dimension, a concept that captures the notion of size of the omega-language satisfying a linear temporal logic formula. We identify the conditions under which this measure is guaranteed to distinguish between weaker and stronger GR(1) formulae. We evaluate our proposed weakness measure in the context of computing GR(1) assumptions refinements.
\end{abstract}

\section{Introduction}

Specifications provide significant aid in the formal analysis of software supporting tasks such as  their verification  and implementation.
However writing such specifications is difficult and error-prone, often resulting  in their incompleteness, inconsistency and unrealizability \cite{Konighofer2009}. Hence providing formal and rigorous support for ensuring their highest quality is of key importance \cite{Kupferman:2012}.
One crucial quality metric for specifications, which this paper focuses on,  is that of weakness in the context of   reactive synthesis \cite{Albarghouthi:2016,Alur2015,Cavezza2017,DIppolito:2015a}.

Reactive synthesis is concerned with finding a system implementation that satisfies a given specification under all possible environments \cite{Pnueli:1989}. 
When no such implementation exists, a specification is said to be unrealizable \cite{Cimatti2008a}. Though there may be many reasons for why a specification is unrealizable, a common  cause  is an  incomplete set of assumptions over the environment behaviour.  
Several techniques \cite{Alur2013,Alur2015,Cavezza2017,Li2011a} have been proposed in order to compute refinements for
incomplete assumptions so as to ensure the realizability of a specification. These approaches consider
specifications expressed in 
a subset of linear temporal logic (LTL), namely generalized reactivity of rank 1 (GR(1)) \cite{Bloem2007,Bloem2012,Braberman2013a}, for which 
tractable synthesis methods exist. Their aim is to find the ``weakest'' assumptions amongst possible  alternatives. 

\emph{Assumption Weakness} \cite{Seshia2015} is a feature intended to capture the degree of freedom (or permissiveness) an environment satisfying the assumptions
has over its behaviours; generally, weaker assumptions are preferred since they allow for more general solutions to the synthesis problem \cite{Chatterjee2008,Seshia2015}.
%
%
%
Existing approaches formalize the weakness relation between assumptions through logical implication \cite{Alur2013,Seshia2015}, i.e., a formula $\phi_1$ is weaker than a formula $\phi_2$ if $\phi_2 \rightarrow \phi_1$ is valid. However, this notion does not fully capture the weakness concept as permissiveness \cite{discrete:2008}. Consider the simple example of a bus arbiter whose environment consists of three devices that can request for bus access. 
Let $r_i$ be the binary signal meaning ``device $i$ requests access''.
An assumption like ``device $1$ requests access infinitely often'' ($\always\eventually r_1$ in LTL) is intuitively less constraining than ``device $2$ and $3$ request access infinitely often'' ($\always\eventually (r_2 \land r_3) $). However, since the two assumptions refer to disjoint subsets of variables, no implication relation holds between the two. 


To enable comparison between weakness of specifications as in the case above, 
we propose a quantitative measure for the weakness of  GR(1) formulae---based on their interpretation as an \olanguage---and a procedure to compute it. 
The measure builds upon the notion of  Hausdorff dimension \cite{Staiger:2015}, a quantity providing an indication of the size of an \olanguage: the higher the dimension, the wider the collection of distinct \owords contained in the \olanguage. 
We show that a sufficient condition for assumptions expressed as invariants to be comparable through our measure is the \textit{strong connectedness} of the underlying \olanguage. To compare assumptions containing fairness conditions, we identify and measure a language decomposition based on fairness complements. Though we focus on comparing the weakness of assumptions refinements, the applied scope of  our weakness metric  can be extended to other contexts, e.g., quantitative model checking, in the form of a measure of the set of behaviors violating some given property (see \cite{Asarin:2014}) and specification coverage as in \cite{Barnat:2016,Tan:2004}.


The paper is structured as follows. Related work is  presented in Sec.~\ref{sec:related}. Notation and background concepts are presented in Sec.~\ref{sec:Background}. In Sec.~\ref{sec:WeaknessHausDim} we define requirements on a weakness measure in an axiomatic form. In Sec.~\ref{sec:WeaknessGR1}, we define Hausdorff dimension and explore its relationship with weakness; hence we introduce the proposed weakness measure first for simpler then for generic GR(1) formulae, and provide sufficient conditions guaranteeing its consistency with implication. We also present our refinement of Staiger's algorithm to compute the weakness measure in the GR(1) case. Sec.~\ref{sec:CaseStudies} presents several applications of our weakness measure to existing GR(1) benchmarks. Omitted details of the experiments and the source code are also available online in \cite{Davide}. Finally, conclusions are drawn in Sec.~\ref{sec:Conclusion}. Some proofs are relegated to the appendix.

\section{Related Work}
\label{sec:related}
\vspace{-0.1cm}
The closest notion to our measure is the \emph{entropy} of \olanguages applied by Asarin et al.~\cite{Asarin:2014,Asarin:2014a} to quantitative model checking. This quantity measures how diverse  the \owords contained in the language of an LTL formula are. However, it is not sufficiently fine-grained to distinguish between weaker and stronger fairness conditions \cite{Asarin:2014}. We will show that our metric based on Hausdorff dimension is capable of  making this distinction.


Quality of LTL formulae has also been defined in the context of model verification.
The work by Henzinger et al.~\cite{Henzinger2010,Henzinger2013} defines a similarity measure between models of LTL formulae so as to render the model checking output quantitative: instead of returning a true/false response, quantitative model checking computes the distance (\emph{stability radius}) of the model from the boundary of the satisfiability region of an LTL property. The scope of our work is different: the measure we propose can be interpreted as the \emph{extension} of such a satisfiability region, which is independent of a specific model to check against.

An alternative way to measure behaviour sets is via probabilities. Probabilistic model checking \cite{Hansson1994,Kwiatkowska2007} enhances the syntax and semantics of temporal logics (usually CTL, \emph{computation tree logic}) with probabilities. This allows for the expressions of properties like ``the probability of satisfying a temporal logic formula $\phi$ by the modelled behaviours is at most $p$.''
Further extensions of LTL and/or automata with preference metrics alternative to probabilities have been proposed in \cite{Almagor2013,Bloem2009,Chatterjee2006,Chatterjee2008}. The difference between using probabilities/preference metrics and our proposal is that while all of these measures are additional and depend on arbitrary parameters that may not reflect the true weakness of a logical formula, the measure we propose quantifies a concept of weakness \emph{intrinsic} to the LTL formula itself.

The problem of identifying weakest assumptions appears in the context of assume-guarantee reasoning \cite{Cobleigh2003a,Lomuscio2010,Nam2006} for compositional model checking. In order to perform model checking of large systems, those systems are generally broken down to components that can be checked independently for correctness. In this context, one of the challenges is to identify the most general (weakest) assumptions over the environment in which each component operates, such that when they are satisfied, the correctness of the entire system is guaranteed. Assumptions are formalized as transition systems (e.g., modal transition systems) rather than declarative LTL specifications, which is the focus of our work.

\section{Preliminaries}
\label{sec:Background}

\textbf{Languages and Automata.}
%
Let $\Sigma$ be a finite set of symbols, which we call \emph{alphabet}. A \emph{word} over $\Sigma$ is a finite sequence of symbols in $\Sigma$. An \emph{\oword} is an infinite sequence of such symbols. A set of words is called a \emph{language}, while a set of \owords is called an \olanguage. A word $w$ is explicitly denoted as a sequence of its symbols $w_1w_2 \dots w_n$, or with a parenthesis notation $(w_1,w_2,\dots,w_n)$, with the symbols separated by commas; the same notation is used for \owords. The notation $w^j$ denotes the suffix of $w$ starting with $w_j$.

Given two words $v$ and $w$, their concatenation is denoted as $v \cdot w$ or simply $vw$. The same notation is used for the concatenation of a word $v$ and an \oword $w$; the concatenation of an \oword and a word is not defined.
Given a set $V$ of finite-length words and a set $W$ of finite-length words or \owords over the same alphabet $\Sigma$, the set $V \cdot W$ is the set of words obtained by concatenating a word in $V$ with a word in $W$. \emph{Kleene's star operator} yields the set $V^*$ of finite words obtained by concatenating an arbitrary number of words in $V$. The \emph{omega operator} applied to $V$ yields the set $V^\omega$ of \owords obtained by concatenating a (countably)  infinite number of words in $V$. Naturally, $\sigmastar$ and $\sigmaomega$ represent, respectively, the set of all finite words and all \owords over the alphabet $\Sigma$. The star and omega operators can also be applied to single finite-length words, like in $w^*$ and $w^\omega$. 

Given an \olanguage $L \subseteq \sigmaomega$, we denote by $\prefnlanguage{n}{L}$ the set of all $w \in \sigmastar$ such that $w$ is a prefix of a word in $L$ and $|w| = n$. We also define $\preflanguage{L} = \bigcup_{n \in \mathbb{N}} \prefnlanguage{n}{L}$ the set of all the prefixes of \owords in $L$.
It is possible to define a topology on $\sigmaomega$. For more details, we refer the reader to \cite{Staiger:2015}. In this context, we only need the notions of closed \olanguages and of their closure. An \olanguage $L$ is \emph{closed} if and only if for any \oword $w$ such that $\preflanguage{\{w\}} \subseteq \preflanguage{L}$, $w \in L$. In other words, $L$ is closed if whenever a word $w$ is arbitrarily close (up to a prefix of arbitrary length) to some word in $L$, then $w \in L$. The \emph{closure} of an \olanguage $L$, denoted by $\closure(L)$, is the smallest closed \olanguage that contains $L$.

The notion of regular \olanguages encompasses \olanguages that allow a finite representation through automata.
Formally, we define a \emph{regular \olanguage} as an \olanguage which is accepted by a deterministic Muller automaton. A \emph{deterministic Muller automaton} (DMA) is defined by the quintuple $\mautomaton = \left\langle Q,\Sigma,q_0,\delta,T \right\rangle$, where $Q$ is a set of states, $\Sigma$ is the alphabet of the \olanguage, $q_0$ is the initial state, $\delta: Q \times \Sigma \rightarrow Q$ is the transition (partial) function and $T \subseteq 2^Q$ is a set (a table) of accepting state sets.
Given an \oword $w \in \sigmaomega$, the \emph{run} induced by $w$ onto $\mautomaton$ is a sequence of states $\mautomaton(w) = q_0q_1\dots$ such that $q_0$ is the initial state and $q_i = \delta(q_{i-1},w_i) \: \forall i \in \mathbb{N}$. Let $\Inf(w) \subseteq Q$ be the set of states occurring infinitely many times in $\mautomaton(w)$. Then an \oword is said to be \emph{accepted} by $\mautomaton$ iff $\Inf(w) \in T$. By extension, the \olanguage accepted by $\mautomaton$ is the set of \owords accepted by $\mautomaton$.

A \emph{deterministic B\"uchi automaton} (DBA) $\bautomaton$ is defined in the same way as a DMA except for the acceptance condition, which is stated in terms of a subset of states $F \subseteq Q$.  A word $w$ is accepted by $\bautomaton$ iff $\Inf(w) \cap F \neq \varnothing$.
Given a DBA it is always possible to obtain an equivalent DMA by replacing the B\"uchi acceptance condition with the table $T=\{Q' \in 2^Q \text{ } | \text{ } Q' \cap F \ne \varnothing \}$. In Sec.~\ref{sec:CaseStudies} we also refer to nondeterministic automata, where the transition function is replaced by a transition relation and the initial state by a set of initial states.

\vspace{6pt}
\noindent
\textbf{Linear Temporal Logic and GR(1).}
\emph{Linear temporal logic} (LTL) \cite{Pnueli:1977} is an extension of Boolean logic with temporal operators. It allows for expressing properties of infinite sequences of assignments to a set $\var$ of Boolean variables. Details of its syntax and semantics are given in Appendix~\ref{app:syntax} for completeness.

In this paper, we deal with a specific subset of LTL, called \emph{Generalized Reactivity (1)} (GR(1)), which is largely employed in controller synthesis \cite{Bloem2010,Bloem2012,Konighofer2009}.  
This subset makes use of the operators $\always$ (``always''), which states that its operand formula must hold at each step of a valuation sequence, $\eventually$ (``eventually''), which requires its operand formula to hold at some point in the sequence, and $\next$ (``next''), which states that the operand formula must hold in the state following the one on which the formula is evaluated.

A GR(1) formula over a set of variables $\var$ has the form
$\phi = \phi^\env \rightarrow \phi^\sys$, 
where $\phi^\env$ and $\phi^\sys$ are conjunctions of the following units: $(i)$ an \emph{initial condition}, which is a pure Boolean expression over variables in $\var$, denoted by $\boolexpr{\init}{}{\var}$; $(ii)$ one or more \emph{invariants}, conditions of the form $\always \boolexpr{\inv}{}{\var\cup\next\var}$, where $\boolexpr{inv}{}{\var\cup\next\var}$ denotes a pure Boolean expression over the set of variables in $\var$ and the set of atoms obtained by prepending an $\next$ operator to each variable; and $(iii)$ one or more \emph{fairness conditions} of the form $\always\eventually \boolexpr{\fair}{}{\var}$.

The semantics of GR(1), as of LTL, 
are formalized as \emph{\owords} over the alphabet $\Sigma = 2^\var$. The set of \owords that satisfy a  formula $\phi$ is a regular \olanguage \cite{Vardi:1996} denoted by $L(\phi)$. 

\section{Problem Statement}
\label{sec:WeaknessHausDim}
In this section, we present an axiomatization of weakness of an LTL formula. Hereafter,  we denote the weakness measure of the LTL formula $\phi$ as  $d(\phi)$: the higher this measure, the weaker $\phi$ is, i.e., $\phi_2$ is weaker than $\phi_1$ if $d(\phi_1) \le d(\phi_2)$.
 
In settings such as \cite{Albarghouthi:2016,Alur2013,Seshia2015}, an LTL formula $\phi_2$ is \emph{weaker} than $\phi_1$ if and only if $\phi_1 \rightarrow \phi_2$ is valid (that is, it is true for any \oword). Semantically, this translates to language inclusion: namely, $\phi_2$ is weaker than $\phi_1$ iff $L(\phi_1) \subseteq L(\phi_2)$. This gives us the first axiom of weakness.
\begin{axiom}\label{axiom1}
	Given two LTL formulae $\phi_1$ and $\phi_2$, if $\phi_1 \rightarrow \phi_2$, then $d(\phi_1) \leq d(\phi_2)$.
\end{axiom}
Notice that this criterion defines a partial ordering of specifications: if none of the two formulae implies the other, those are incomparable according to this criterion.
However, even for the incomparable case it may be useful to define a preference criterion. 

Consider the simple case of two invariants over $\var = \{a,b,c\}$, $\phi_1 = \always (a \land b)$ and $\phi_2 = \always c$. Even if the two formulae are incomparable according to implication, i.e., neither one implies the other, it is clear that $\phi_1$ allows in some sense fewer behaviors than $\phi_2$: at each time step, the former allows for 2 distinct valuations of $\var$ while $\phi_2$ allows 4 of them.

Consider  the formulae 
$\phi_3 = \always(a \rightarrow \next b)$ and $\phi_4 = \always((a \land b) \rightarrow \next c)$ instead. Despite neither implying the other, we note that $\phi_3$ is more restrictive than $\phi_4$ asymptotically: that is, for a large enough $n$, the number of finite prefixes of length $n$ that satisfy $\phi_3$ is less than the number of finite prefixes of length $n$ satisfying $\phi_4$ ($\setsize{L(\phi_3)} < \setsize{L(\phi_4)}$). This can be easily understood if one considers that $\phi_3$ poses a restriction to the next symbol in an \oword whenever $a$ is true (which holds in $4$ out of $8$ possible valuations of $\var$), while $\phi_4$ poses a similar restriction when $a \land b$ holds (in $2$ out of the $8$ valuations).

This means that weakness of a formula should be formalized, in addition to Axiom \ref{axiom1},  in terms of the number of finite prefixes it allows.
 Formally:
\begin{axiom}\label{axiom2}
	Given two LTL formulae $\phi_1$ and $\phi_2$,  $\phi_2$ is said to be weaker than $\phi_1$ if there exists some length $\bar{n}$ such that, for every $n > \bar{n}$, the set of prefixes of length $n$ in $L(\phi_2)$ contains more elements than the set of prefixes of the same length in $L(\phi_1)$, i.e., if $\forall{n}>\bar{n},\; \setsize{\prefnlanguage{n}{L(\phi_2)}} \geq \setsize{\prefnlanguage{n}{L(\phi_1)}}$, then $d(\phi_1) \leq d(\phi_2)$.
\end{axiom}


The final desirable property is that a weakness measure be at least as discriminating as implication in case one formula strictly implies the other.
\begin{axiom}\label{axiom3}
	Let $\phi_1$ and $\phi_2$ be such that $\phi_1 \rightarrow \phi_2$ is valid and $\phi_2 \rightarrow \phi_1$ is not. Then $d(\phi_1) < d(\phi_2)$.
\end{axiom}
In the next section, we prove that our proposed  weakness measure satisfies Axioms
\ref{axiom1} and \ref{axiom2}. 
We then show that, although our weakness measure is not guaranteed to satisfy Axiom \ref{axiom3} in general, we are able to guarantee  so 
for a specific class of formulae.

\section{Weakness Measure of GR(1) Formulae}
\label{sec:WeaknessGR1}

\textit{Hausdorff dimension} and \textit{Hausdorff measure} are basic concepts in fractal geometry and represent a way to define measures of extension---that is, analogous concepts to length, area, volume from classical geometry---for fractals \cite{Falconer:2004}. Staiger \cite{Staiger:2015} pinpointed a homeomorphism between fractals and regular \olanguages and proposed an analogous interpretation of the two quantities as extension measures of \olanguages.
Intuitively, given an \olanguage $L$, its Hausdorff dimension quantifies the growth rate of the number of distinct $n$-long prefixes of words in the language, over the length $n$ of those prefixes. This makes it a good candidate for quantifying weakness: the less constrained the language is, the more prefixes of a fixed length are contained in it,  implying a higher Hausdorff dimension.

The formal definition of Hausdorff dimension is tightly related to the notion of Hausdorff measure. The following definitions are given in \cite{Staiger:1998}.
\begin{definition}
	[$\alpha$-dimensional Hausdorff outer measure]
	\label{def:HausAlphaMeas}
	Given a regular \olanguage $L$ over an alphabet $\Sigma$ with cardinality $r$, and a nonnegative real value $\alpha$, the {\em $\alpha$-dimensional Hausdorff outer measure} of L is defined as
	\begin{equation}
	\label{eq:outerMeasure}
	\Hausmeas{\alpha}{L} = \lim_{n \rightarrow \infty} \inf_{V \in \mathcal{L}_n} \sum_{v \in V}r^{-\alpha|v|}
	\end{equation}
	where $\mathcal{L}_n = \left\{V \subseteq \sigmastar \mid V \cdot \sigmaomega \supseteq L \text{ and } |v| \geq n \text{ for all } v \in V \right\}$ is the collection of languages $V$ containing finite words of length at least $n$ and such that every word in $L$ has at least a prefix in $V$. \qed
\end{definition}

\begin{definition}
	[Hausdorff dimension and measure]
	\label{def:HausDimMeas}
	Given an \olanguage $L$, its {\em Hausdorff dimension}, denoted by $\Hausdim{L}$, is the (unique) value $\bar{\alpha}$ such that
	$$\Hausmeas{\alpha}{L}=\infty \;\; \alpha < \bar{\alpha}$$
	$$\Hausmeas{\alpha}{L}=0 \;\; \alpha > \bar{\alpha}$$
	The value $\Hausmeas{\Hausdim{L}}{L}$ is called the {\em Hausdorff measure} of $L$. \qed
\end{definition}

In other words, Hausdorff measure is the limit of the process of approximating the \olanguage $L$ by a set $V$ of finite prefixes with length at least $n$, and weighing each prefix with a quantity $r^{-\alpha|v|}$ that decreases as the prefix length increases. This limit can be finite and positive for at most one value of the $\alpha$ parameter. This value is  called \emph{Hausdorff dimension}.

A related concept appearing in the literature is entropy:
\begin{definition}[Entropy \cite{Merzenich:1994}]\label{def:entropy}
	Given an \olanguage $L \subseteq \sigmaomega$ over an alphabet of size $r$, the \emph{entropy} of $L$ is $\entropy{L}=\limsup_{n\rightarrow\infty} \frac{1}{n} \log_r\setsize{\prefnlanguage{n}{L}}\:.$
\end{definition}
It has been proved \cite{Merzenich:1994} that the Hausdorff dimension has a close relationship with the notion of entropy:
Specifically, we have $\Hausdim{L} \leq \entropy{L}$ in general, and $\Hausdim{L}=\entropy{L}$ if $L$ is a closed \olanguage. 
Details on how entropy is computed are given in Appendix~\ref{app:entropy}.

When $L$ is not closed, the general algorithm presented in \cite{Staiger:1998,Staiger:2015} provides a more refined intuition of what is actually quantified by Hausdorff dimension, which distinguishes it from entropy. The algorithm is based on computing a Muller automaton $\mautomaton_L$ accepting $L$ with set of accepting state sets $T_L$. For each accepting set $S' \in T_L$ and for each state $s \in S'$, consider the \olanguage $C_{S'}$ consisting of all the infinite paths in $\mautomaton_L$ starting from $s$ and visiting no states outside $S'$. It can be shown that this language is closed and its entropy $\entropy{C_{S'}}$ is independent of the choice of $s$ \cite{Staiger:1998}. The Hausdorff dimension of $L$ is then
\begin{equation}
\label{eqn:HausdimMuller}
\Hausdim{L} = \max_{S' \in T_L} \entropy{C_{S'}} \:.
\end{equation}

Hausdorff dimension provides an ordering consistent with the weakness notion defined in Sec.~\ref{sec:WeaknessHausDim}.
We can interpret it as a measure of the asymptotic degrees of freedom of an \olanguage: it quantifies how many different evolutions are allowed to an \oword once its run remains in an accepting subset of the Muller automaton. The example below shows how it differs from entropy.
\begin{wrapfigure}[6]{r}{4cm}
	\begin{center}
		\vspace{-1cm}
		\includegraphics[width=0.7\linewidth]{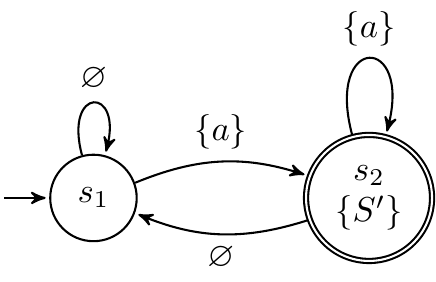}
		\caption{DMA of $L(\phi_1)$.}
		\label{fig:MullerPhi1EntropyHausdim}
	\end{center}
	\vspace{-10cm}
\end{wrapfigure}

\begin{example}
	\label{ex:FGFormulaEntropyHausdim}
	Consider the LTL formula $\phi_1 = \eventually \always a$ over the variable set $\var = \{a\}$ whose Muller automaton is shown in Fig.~\ref{fig:MullerPhi1EntropyHausdim}.  The collection of accepting sets to which a state belongs is enclosed in curly braces.

	Notice that for any $w \in L(\phi_1)$ both valuations of $\var$ are allowed until $w$ reaches the accepting state, and the satisfaction of $\always a$ may be delayed arbitrarily. Therefore, for any finite $n$, $\setsize{\prefnlanguage{n}{L}} = 2^n$, and thereby $\entropy{L(\phi_1)} = 1$.
	
	In this simple DMA, there is only one accepting singleton $\{s_2\}$. Therefore, there is only one $C_{S'} = \{\{a\}^\omega\}$ which allows only the symbol $\{a\} \in 2^\var$. This implies $\setsize{\prefnlanguage{n}{C_{S'}}} = 1$. The Hausdorff dimension is $\Hausdim{L(\phi_1)} = \entropy{C_{S'}} = 0$. 
	This example demonstrates that the Hausdorff dimension isolates the asymptotic behaviour of $L(\phi_1)$ as it depends only on the condition $\always a$ that is eventually satisfied by any \oword in the \olanguage. \qed
\end{example}

The following theorem  shows that Hausdorff dimension is consistent with implication (hence satisfying Axiom \ref{axiom1}).

\begin{theorem}
	\label{thm:HausDimAndWeakness}
	Given two LTL formulae $\phi_1$ and $\phi_2$ such that $\phi_1 \rightarrow \phi_2$ is valid, $\Hausdim{L(\phi_1)} \leq \Hausdim{L(\phi_2)}$.
	\proof \emph{
	This follows from the language inclusion $L(\phi_1) \subseteq L(\phi_2)$ and the monotonicity of Hausdorff dimension with respect to language inclusion \cite{Merzenich:1994}.}
\end{theorem}

Note that Theorem~\ref{thm:HausDimAndWeakness} does not exclude the situation where one formula strictly implies another, but the two languages have the same Hausdorff dimension, thus violating Axiom~\ref{axiom3}. 
We investigate under which conditions this holds in the context of GR(1) formulae and provide a refined 
weakness measure 
that bounds the number of cases in which it can happen.

To this end,  in what follows, we introduce a new weakness measure for GR(1) based on Hausdorff dimension. We first analyse the dimension of invariants. We then show that under the condition of strong connectedness, it is possible to distinguish between weaker and stronger invariants, in the implication sense (Sec.~\ref{sec:DimInvariants}). 
We show how, under the same condition, this measure fails to capture the impact of conjoining a fairness condition (Sec.~\ref{sec:Fairness}). 
To overcome this, we define a refined weakness measure for GR(1) formulae that comprises two components:  the Hausdorff dimension $(i)$ of the whole formula and  $(ii)$ of the difference language between the invariant and the fairness conditions (Sec.~\ref{sec:DimPairs}).

\subsection{Dimension of Invariants}
\label{sec:DimInvariants}
Consider the formula $\invariant = \always \boolexpr{}{}{\var \cup \next\var}$. The \olanguage $L(\invariant)$ is closed. Hence, the Hausdorff dimension of $L(\phi^{\inv})$ coincides with its entropy $\entropy{L(\invariant)}$ and can be computed as the maximum eigenvalue of the adjacency matrix of its B\"uchi automaton (see Appendix~\ref{app:entropy}). From this equivalence and Definition \ref{def:entropy}, it is easy to see that in this case Hausdorff dimension satisfies Axiom~\ref{axiom2}.
\begin{wrapfigure}[11]{r}{4cm}
\vspace{-0.3cm}
		\centering
		\subfloat{\includegraphics[width=0.2\linewidth]{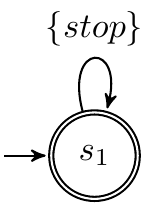}}$\:\:$\\
		\subfloat{\includegraphics[width=0.7\linewidth]{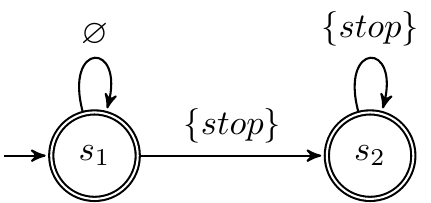}}
		\caption{DBAs of $\invariant_1$ (top) and $\invariant_2$ (bottom)}
		\label{fig:BuchiStopNotStop}
\end{wrapfigure}
In general, Theorem~\ref{thm:HausDimAndWeakness} may hold for invariants where one is strictly weaker than the other and both have equal dimensions
as demonstrated in the following.

\begin{example}
	\label{ex:NonStronglyConnected}
	Consider the variable set $\var = \{\stopp\}$ and the formulae $\invariant_1 = \always \stopp$ and $\invariant_2 = \always(\stopp \rightarrow \next \stopp)$. Their B\"uchi automata are shown in Fig.~\ref{fig:BuchiStopNotStop}. Clearly $\invariant_1 \rightarrow \invariant_2$ strictly, however the two languages have the same Hausdorff dimension $\Hausdim{L(\invariant_1)} = \Hausdim{L(\invariant_2)} = 0$.
\end{example}

There exists, however, a subclass of invariants for which the dimension is strictly monotonic with respect to  implication. 
This subclass is characterized through the concept of \emph{strong connectedness} of an \olanguage. 
Hereafter, given a word $w \in \preflanguage{L}$, we denote by $\suffixlanguage{w}{L}$ the \olanguage formed by the \owords $v$ such that $wv \in L$ (that is, the suffixes allowed in $L$ after reading $w$).

\begin{definition}[Strongly connected \olanguage \cite{Merzenich:1994}]
\label{def:StronglyConnectedOlanguage}
	An \olanguage $L$ is {\em strongly connected} if for every prefix $w \in \preflanguage{L}$ there exists a finite word $v \in \sigmastar$ such that $\suffixlanguage{wv}{L} = L$.
\end{definition}
In other words, an \olanguage is strongly connected if and only if there exists a strongly connected finite-state automaton which represents it \cite{Merzenich:1994}, i.e.,  an automaton such that given any pair of states, each of them is reachable from the other.
%
%
Using this notion, in the next theorem we provide a sufficient condition over invariants for Axiom~\ref{axiom3} to be satisfied (the proof is relegated to Appendix~\ref{app:ThmProof}):
\begin{theorem}
\label{thm:InvariantStrictWeakness}
	Let $\invariant_1 = \always B_1(\var \cup \next\var)$ and $\invariant_2 = \always B_2(\var \cup \next\var)$ be two non-empty invariants such that $\invariant_1 \rightarrow \invariant_2$ is valid, $\invariant_2 \rightarrow \invariant_1$ is not valid and $\invariant_2$ is strongly connected. Then $\Hausdim{L(\phi^{inv}_1)}<\Hausdim{L(\phi^{inv}_2)}$.
\end{theorem}

\begin{wrapfigure}[7]{r}{3.5cm}
\vspace{-0.5cm}
		\begin{center}
		\includegraphics[width=0.2\linewidth]{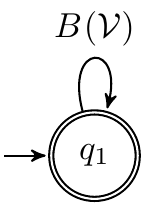}
		\caption{DBA of a one-state invariant. }
		\label{fig:BuchiOneStateInvariant}
		\end{center}
\end{wrapfigure}
	An interesting kind of invariant that falls in this class is the \emph{one-state invariant}, one that does not use the $\next$ operator: $\invariant_s = \always \boolexprvar$ whose DBA  is shown in Fig.~\ref{fig:BuchiOneStateInvariant}. (For succinctness, the set of valuations that label a transition between the same states is denoted by the Boolean expression characterizing it.)
%
	In this case, the Hausdorff dimension has a closed form:
	$$\Hausdim{\invariant_s} = \log_r \setsize{\boolexprvar}$$
	where $r=2^\setsize{\var}$ is the number of valuations of $\var$ and $\setsize{\boolexprvar}$ is the number of valuations that satisfy $\boolexprvar$. Invariants of this type are clearly strongly connected and satisfy Theorem~\ref{thm:InvariantStrictWeakness}.

\begin{remark}
	Typical examples of GR(1) specifications manually produced, like those of device communication protocols, make use of strongly connected environment assumptions. It is indeed natural to allow environments to be reset to their initial state after some steps. However, when specifications contain ``until'' operators or response patterns, the procedure to convert them into GR(1) \cite{Maoz2015} may yield assumptions which are no longer strongly connected. In those cases, a problem similar to that of Example~\ref{ex:NonStronglyConnected} may arise. \qed
\end{remark}

\subsection{Fairness and Fairness Complements}
\label{sec:Fairness}
\begin{wrapfigure}[12]{r}{4.3cm}
\vspace{-1cm}
		\begin{center}
	\includegraphics[width=0.6\linewidth]{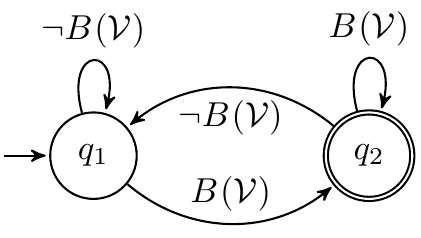}\\[1ex]
	\includegraphics[width=0.6\linewidth]{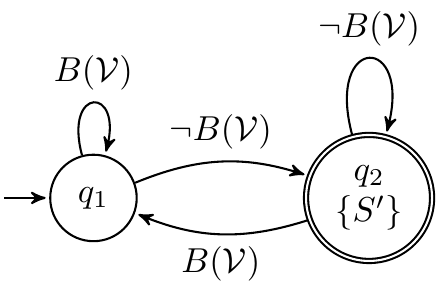}
	\caption{DBA of $L(\fairness)$ (top) and DMA of $L(\cfairness)$ (bottom).}
	\label{fig:BuchiFairness}
		\end{center}
\end{wrapfigure}

Consider the generic fairness condition $\fairness = \always\eventually \boolexprvar$ whose DBA is shown in Fig.~\ref{fig:BuchiFairness}. 
 This language is not closed: take a symbol $x \in \Sigma$ that does not satisfy $\boolexprvar$ and the \oword $x^\omega$. It is clear that $\preflanguage{\{x^\omega\}} \subseteq \preflanguage{L(\fairness)}$, but $x^\omega \not\in L(\fairness)$. 
We apply the algorithm in Sec.~\ref{sec:WeaknessGR1} (cf. equation \ref{eqn:HausdimMuller})
for non-closed languages. A DMA for $L(\fairness)$ can be obtained from the top DBA in Fig.~\ref{fig:BuchiFairness}: the accepting sets are $S'_1 = \{q_1,q_2\}$ and $S'_2 = \{q_2\}$. It is easy to see that $\entropy{C_{S'_1}}=1$ and $\entropy{C_{S'_2}}=\log_r\setsize{\boolexprvar}\le 1$. Therefore, $\Hausdim{L(\fairness)}=1$, independently of $\boolexprvar$. 
We conclude that fairness conditions are  indistinguishable from the $\true$ constant, which also has dimension $1$.
%

%
%
To allow for a distinction to be made, we  characterize the negation of such formula. We call an LTL formula of the kind $\cfairness = \eventually \always \lnot \boolexprvar$ a \emph{fairness complement}. The DMA of $L(\cfairness)$ is shown in the bottom of Fig.~\ref{fig:BuchiFairness}.
The only accepting set is $S' = \{q_2\}$. (Notice that unlike the top one, this automaton accepts only words that stay forever in $q_2$ from a certain step on.) 
The language $C_{S'}$ (see Sec.~\ref{sec:WeaknessGR1}) has an entropy of $\log_r\setsize{\lnot \boolexprvar}$. Hence
$$\Hausdim{L(\cfairness)} = \log_r\setsize{\lnot \boolexprvar} $$
where $r = 2^\setsize{\var}$.
Notice that $C_{S'}$ is the language of the formula $\always \lnot \boolexprvar$, which is an ``asymptotic'' condition of $\cfairness$. As 
observed previously, Hausdorff dimension is strictly monotonic for one-state invariants. Therefore, the
weakness of fairness complements can be ranked in terms of the Hausdorff dimension, allowing to compare fairness conditions as follows:
\begin{theorem}
	Let $\fairness_1$ and $\fairness_2$ be two fairness conditions such that $\fairness_1 \rightarrow \fairness_2$ is valid and $\fairness_2 \rightarrow \fairness_1$ is not valid. Then $\Hausdim{L(\lnot\fairness_1)} > \Hausdim{L(\lnot\fairness_2)}$.
\end{theorem}
In other words, the stronger a fairness formula is, the weaker its complement and thereby the higher its dimension.

\subsection{Dimension Pairs for GR(1) Formulae}
\label{sec:DimPairs}

 \begin{wrapfigure}[19]{r}{4cm}
	\centering
	\includegraphics[width=.8\linewidth]{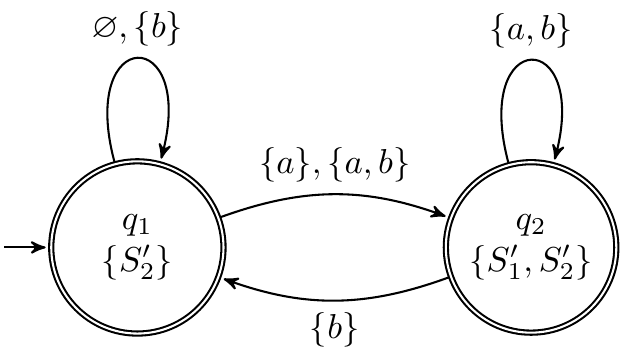}\\
	\vspace{10pt}
	\includegraphics[width=.8\linewidth]{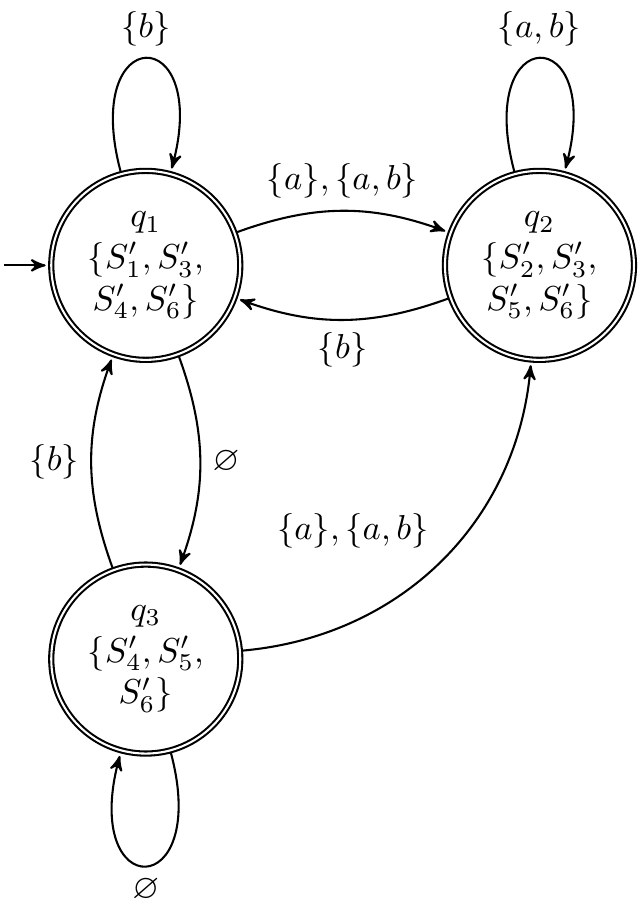}
	\vspace{-5pt}
	\caption{DMAs of $\phi_1$ (top) and $\phi_2$ (bottom) of Example~\ref{ex:GR1Fairness}.}
	\label{fig:ExampleGR1Fairness}
\end{wrapfigure}
Consider a generic GR(1) formula $\phi = \initial \land \invariant \land \bigwedge_{i=1}^m \fairness_i$. 
We show through an example that even when $\invariant$ is strongly connected, Hausdorff dimension may not distinguish between weaker and stronger fairness conditions in the implication sense. This problem has been previously pointed out in the work of \cite{Asarin:2014}.

\begin{example}
\label{ex:GR1Fairness}
	Consider the two formulae over the variables $\var = \{a,b\}$: $\phi_1 = \always(a \rightarrow \next b) \land \always\eventually a$ and 
	$\phi_2 = \always(a \rightarrow \next b) \land \always\eventually b$.
	The same invariant appears in both, and thereby have the same Hausdorff dimension, but the fairness condition in $\phi_2$ is always satisfied when the fairness condition of $\phi_1$ is satisfied, by virtue of the invariant itself. However, the \oword $\{b\}^\omega$ satisfies $\phi_2$ but not $\phi_1$. So, $\phi_1$ implies $\phi_2$ but not vice versa.
 
	The language of both formulae is not closed. The Muller automata of $\phi_1$ and $\phi_2$ are shown at the top and bottom, respectively, in Fig.~\ref{fig:ExampleGR1Fairness}. 
	In both automata, there is an accepting set that covers the entire state space ($S'_2$ in $L(\phi_1)$ and $S'_6$ in $L(\phi_2)$. It is possible to show that the maximum $\entropy{C_{S'}}$ of equation (\ref{eqn:HausdimMuller}) is achieved exactly for these accepting sets \cite{Berman:1994,Merzenich:1994}. The \olanguages $C_{S'_2}$ in $L(\phi_1)$ and $C_{S'_6}$ in $L(\phi_2)$ both coincide with the language of the invariant alone. Therefore,
$$\Hausdim{\phi_1} = \Hausdim{\phi_2} = \Hausdim{L(\always(a \rightarrow \next b))}\:.$$
\end{example}

To distinguish between the two formulae, we exploit the fact that the complement of a fairness condition is a formula of the kind $\eventually\always \boolexprvar$ which can be compared through Hausdorff dimension. Therefore, we propose a weakness measure which consists of two components: one relating to the whole formula and one measuring the \olanguage excluded from the invariant by the addition of the fairness conditions. 

\begin{definition}[Weakness]
	The \emph{weakness} of a GR(1) formula $\phi=(\initial \land \invariant\bigwedge_{i=1}^m \fairness_{i})$, denoted by $d(\phi)$, is the pair $(d_1(\phi),d_2(\phi))$ such that $d_1(\phi)$ is the Hausdorff dimension of $L(\phi)$; and $d_2(\phi)$ is the Hausdorff dimension of $L(\phi^c) = L(\initial \land \invariant \land \bigvee_{i=1}^m \cfairness_i)$, where $\cfairness_i = \lnot \fairness_i$.
The following partial ordering is defined based on the weakness measure: If $d^i=(d^i_1,d^i_2)$, with $i\in{1,2}$ are weakness measures for two GR(1) formulae, then $d^1<d^2$ if $d^1_1<d^2_1$ or $d^1_1=d^2_1$ and  $d^1_2 > d^2_2$.
\end{definition}
%

\begin{wrapfigure}[17]{r}{4.5cm}
\vspace{-0.2cm}
		\centering
		\includegraphics[width=.6\linewidth]{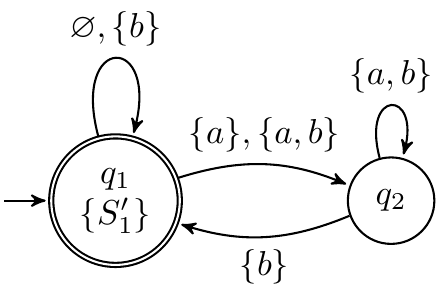}\\
		\vspace{7pt}
		\includegraphics[width=.6\linewidth]{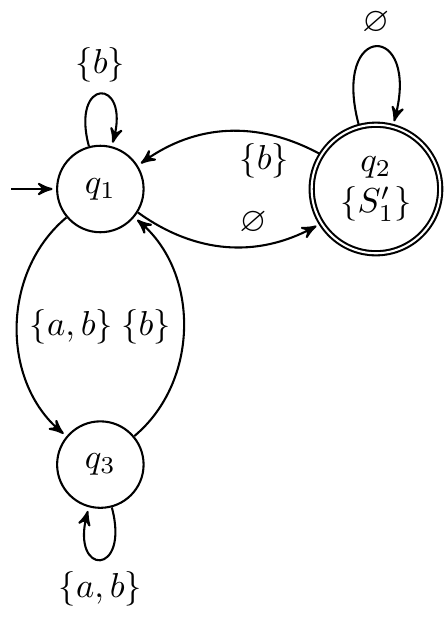}
		\caption{DMAs of  $\phi^c_1$ (top) and  $\phi^c_2$ (bottom) of Example~\ref{ex:WeaknessGR1}.}
		\label{fig:ExampleGR1}
\end{wrapfigure}


We apply below this weakness measure to the formulae in Example~\ref{ex:GR1Fairness}.

\begin{example}
	\label{ex:WeaknessGR1}
	To compute $d_2$, let us define $\phi^c_1 = \always(a \rightarrow \next b)\land \eventually \always \lnot a$ and $\phi^c_2 = \always(a \rightarrow \next b)\land \eventually \always \lnot b$. The DMAs of the resulting languages are shown respectively  in Fig.~\ref{fig:ExampleGR1}.
		%
	Each of them has just one accepting singleton, so the computation of the Hausdorff dimension is straightforward:
	$\Hausdim{\phi^c_1} = \frac{1}{2}$ and $\Hausdim{\phi^c_2} = 0$. In summary, since $\phi_1$ is more restrictive than  $\phi_2$, the Hausdorff dimension of the \olanguage cut out by $\always\eventually a$ is higher than the Hausdorff dimension of the behaviours excluded by $\always\eventually b$.
\end{example}

The following Theorem justifies the use of this dimension pair for weakness quantification when the formulae have the same invariant.
\begin{theorem}
	Let $\phi_1=\invariant \land \bigwedge_{i=1}^m \fairness_{1,i}$ and $\phi_2 = \invariant \land \bigwedge_{j=1}^l \fairness_{2,j}$, such that $\phi_1 \rightarrow \phi_2$ is valid. Then $d_1(\phi_1) = d_1(\phi_2)$ and $d_2(\phi_1) \geq d_2(\phi_2)$.
	\proof \emph{
		Since $\phi_1$ implies $\phi_2$, $L(\phi_1) \subseteq L(\phi_2)$. Furthermore, for $i=1,2$, $L(\phi_i) = L(\invariant) \cap L(\bigwedge_{j=1}^m \fairness_{i,j})$. Therefore, $L(\invariant) \backslash L(\bigwedge_{j=1}^m \fairness_{1,j}) \supseteq L(\invariant) \backslash L(\bigwedge_{j=1}^l \fairness_{2,j})$, i.e., $L(\phi^c_1) \supseteq L(\phi^c_2)$. Then, by monotonicity,  $\dim(\phi^c_1) \ge \dim(\phi^c_2)$, finishing the proof. } \qed
\end{theorem}
Therefore, given two formulae with the same invariant, we deem the formula with lower $d_2$ weaker.

Regarding formulae with the same $d_1$ and different invariants, we justify heuristically the same order relation. We first note that the Hausdorff dimension of a countable union of \olanguages, as noted in \cite{Staiger:2015}, is
$$\Hausdim{\bigcup_{i} L_i} = \sup_i \Hausdim{L_i}.$$
This property is known as the \emph{countable stability} of Hausdorff dimension.
This implies that for any formula $\phi$, if $d_2(\phi) \le d_1(\phi)$ then
$$\Hausdim{L(\invariant)} = \Hausdim{L(\phi) \cup L(\phi^c)} = \Hausdim{L(\phi)} \:.$$
So, if for two formulae, $\phi_1$ and $\phi_2$, we have $d_1(\phi_1)=d_1(\phi_2)>d_2(\phi_1)>d_2(\phi_2)$, then this can be interpreted as the two invariants having the same dimension and the fairness condition of $\phi_1$ removing more behaviours than the fairness condition of $\phi_2$. In this sense, $\phi_2$ is weaker than $\phi_1$. 
This justifies intuitively our weakness definition and the associated partial ordering.
In Sec.~\ref{sec:CaseStudies}, we illustrate applications of this order relation for comparing GR(1) assumptions.

The computation of $d_2(\phi)$ for a generic $\phi$ with $m$ fairness conditions can be reduced to the case of a single fairness condition. Based on the countable stability of Hausdorff dimension, we have
$$d_2(\phi) = \sup_{i=1,\dots,m} d_2(\initial \land \invariant \land \cfairness_i) \:.$$

Furthermore, the case of a single fairness condition can be further reduced to computing the Hausdorff dimension of an invariant by the following theorem.
\begin{theorem}
\label{thm:HausdimComputation}
	Given a formula $\phi^c = \always \boolexpr{inv}{}{\var\cup\next\var} \land \eventually \always \lnot \boolexpr{fair}{}{\var}$ we have
	$$\Hausdim{L(\phi^c)} = \Hausdim{L(\always(\boolexpression^{\text{inv}} \land \lnot \boolexpression^{\text{fair}}))} \:.$$
	
	\noindent \emph Proof sketch (full proof is presented in Appendix~\ref{app:thm_hausdimcomp}). 
	\emph{Since $L(\phi^c)$ is not closed, the Hausdorff dimension must be computed from a DMA. The proof (given in Appendix~\ref{app:thm_hausdimcomp}) consists in showing that the DMA's accepting subsets correspond to the automaton of an \olanguage where both $\boolexpression^{inv}$ and $\lnot \boolexpression^{fair}$ are satisfied at every step.
	This property is a generalization of the observation made in Sec.~\ref{sec:Fairness} about the Hausdorff dimension of fairness complements. } \qed
\end{theorem}

\subsection{Initial Conditions}
\label{sec:InitialConditions}
Consider $\initial = \boolexprvar$. An expression of this form constrains only the first symbol of the \owords in $L(\initial)$. For the same reason as $\fairness$ in Sec.~\ref{sec:Fairness}, $L(\initial)$ is closed, and therefore its dimension can be computed via its entropy. By applying the definition of entropy, it is easy to see that, similarly to  the unconstrained language $L(\true)$, $$\Hausdim{L(\initial)} = 1\:.$$

Consider now a formula $\phi = \initial \land \invariant$. A DBA $\bautomaton$ for $L(\phi)$ can be computed from a DBA $\bautomaton_{inv}$ of $L(\invariant)$ by removing all transitions starting from its initial state whose labels do not satisfy $\boolexprvar$. The resulting automaton may leave out parts of $\bautomaton_{inv}$ that are no longer reachable from the initial state. This does not happen if $L(\invariant)$ is strongly connected, as in that case any non-initial state in $\bautomaton_{inv}$ is reachable from any other state. In this case
$$\Hausdim{\phi} = \Hausdim{\invariant} \:.$$

This implies that the initial conditions do not affect the Hausdorff dimension and hence cannot be always ordered by our weakness measure. This is acceptable since typically, in applications like assumptions refinement, the focus is in assessing invariants or fairness conditions rather than initial conditions \cite{Li2011a}.

\section{Evaluation}
\label{sec:CaseStudies}
We evaluate here our proposed weakness measure through applications to benchmarks  within the assumptions refinement domain, demonstrating its usefulness in distinguishing weakness of different formulae, and discussing the computation time bottlenecks. (In Appendix~\ref{app:Evaluation}, we report on our evaluation within another  application domain, namely quantitative model checking.)

To this aim, we implemented the weakness measure computation  for GR(1) specifications in Python 2.7 and made it publicly available in \cite{Davide}. 
Our implementation makes use of the Spot tool \cite{Duret:2016} for LTL-to-automata conversion.
%
We integrated the weakness computation algorithm within two state-of-the-art counterstrategy-guided assumptions refinement approaches 
 \cite{Alur2013,Cavezza2017} (the implementations are available in \cite{Davide}). The outcome of such approaches is a \emph{refinement tree}, a tree structure where each node is associated with a GR(1) formula consisting of a conjunction of environment assumptions; if we denote by $\phi$ a formula associated with a node, the node's children are of the form $\phi \land \psi$, where $\psi$ is a single initial condition, invariant, or fairness condition. Since the goal of such procedures is identifying weakest formulae that describe an environment, our weakness measure can be used to provide a preference ranking of the tree nodes.

 

We conducted experiments on two benchmarks for GR(1)  assumptions refinement, namely the specifications of a lift controller and of the AMBA-AHB protocol for device communications in its versions for two, four and eight master devices \cite{Alur2013,Bloem2012,Li2011a}. The lift controller example specifies a controller for a lift with three floors: the Boolean variable $b_i$ denotes the state of the button on floor $i$; the Boolean variable $f_i$ is true iff the lift is at floor $i$. For more details on the initial assumptions $\phi^\env$ see \cite{Alur2013}. The AMBA-AHB protocol provides signals for requesting access to a bus ($\hbusreq_i$), for granting access ($\hgrant_i$), for signalling the termination of a communication ($\hready$), and for identifying the current owner of the bus ($\hmaster$). Other signals are detailed in \cite{Bloem2012}. To our knowledge, the AMBA08 specification is one of the biggest benchmarks available in this field, with $55$ binary variables, $28$ initial assumptions and $157$ guarantees.

In the followings we focus examples taken from \cite{Alur2013,Cavezza2017}, and discuss three cases highlighting features of our weakness measure: $(i)$ in the first example, we demonstrate the relationship between weakness and implication; $(ii)$ second, we consider cases when two formulae are not comparable by implication but can be ranked with our measure; and $(iii)$ we discuss the case of formulae equally constraining the environment, which have equal ranking according to our measure. We refer the reader to \cite{Davide} for the complete results.

\vspace{5pt}
\noindent
\textbf{Relation between weakness and implication.} 
 Consider the lift controller example. Two refinements computed by the automated approach in \cite{Cavezza2017} are: $\phi_1 = \always ((\lnot b_1 \land \lnot b_2 \land \lnot b_3) \rightarrow \next (b_1 \lor b_2 \lor b_3))$; and 
$\phi_2 = \always \eventually (b_1 \lor b_2 \lor b_3)$.
The first forces one of the buttons to be pressed at least every second step in a behaviour. The second forces one of the buttons to be pressed infinitely often in a behaviour. It is clear that $\phi_1$ implies $\phi_2$. We compare the assumptions obtained by refining the original assumptions with the first one and with the second one: $d(\phi^\env \land \phi_1) = (0.7746,0)$ and $d(\phi^\env \land \phi_2) = (0.7925,0.5)$.
%
Notice that $d_1(\phi^\env \land \phi_1)<d_1(\phi^\env \land \phi_2)$ and this is consistent with the fact that $\phi_1$ is stronger than $\phi_2$.
Consider now the two fairness refinements: $\phi_2 = \always\eventually(b_1 \lor b_2 \lor b_3)$; and $\phi_3 = \always\eventually b_1$.
%
We have
$d(\phi^\env \land \phi_2) = (0.7925,0.5)$ and  $d(\phi^\env \land \phi_3) = (0.7925,0.695)$.
Here, $d_1$ is equal for both formulae and $d_2(\phi^\env \land \phi_2)<d_2(\phi^\env \land \phi_3)$; this is consistent with the fact that $\phi_2$ is weaker than $\phi_3$.

\vspace{5pt}
\noindent
\textbf{Formulae incomparable via implication.} 
Consider $\phi_3$ above and  $\phi_4 = \always\eventually(b_2 \lor b_3)$.
Neither implies the other. However, it is reasonable to argue that $\phi_4$ is less restrictive than $\phi_3$: while $\phi_3$ constrains exactly one button to be pressed infinitely often, $\phi_4$ allows the extra choice of which one (out of two) . This intuition is indeed reflected  by our computed weakness metric: $d(\phi^\env \land \phi_3) = (0.7925,0.695)$ and  $d(\phi^\env \land \phi_4) = (0.7925,0.5975)$.
This expresses the notion that $\phi_4$ removes less behaviours from $\phi^\env$  than $\phi_3$.

Our weakness measure can help in spotting asymmetries between assumptions that are syntactically equal but constrain semantically different variables. Consider an extended version of the lift controller example including the input variable $\alarm$ and the output variable $\stopp$: whenever $\alarm$ is set to high, the lift enters a $\stopp$ state where it does not move from the floor it is at. The specification of this system is given in the Appendix~\ref{app:ExtLift}. Computing the weakness of the two refinements $\phi_5 = \always \lnot b_1$ and $\phi_6 = \always \lnot \alarm$ yields $d(\phi^\env \land \phi^\sys \land \phi_5) = (0.3694,0.3207)$ and $d(\phi^\env \land \phi^\sys \land \phi_6) = (0.3746,0.3346)$. This is consistent with the intuition that the former assumption excludes a part of the desirable system behaviors (all the ones that allow it to reach floor $1$), while the latter excludes only the error traces ending in the $stop$ state, being then a weaker restriction on the combined behaviors of the controller and the environment.

The following two assumptions refinements are computed 
for the AMBA-AHB case study with two masters: $\psi_1 = \always(\lnot \hbusreq_1 \lor \next(\hready \lor \lnot \hbusreq_1))$; and $\psi_2 = \always((\lnot \hgrant_1 \land \hready \land \hbusreq_1) \rightarrow \next(\lnot \hready \lor \lnot \hbusreq_1))$.
As in the case of the lift example, neither formula implies the other. The weakness of the resulting assumptions is:
$d(\psi^\env \land \psi_1) = (0.9503,0.9068)$ and $d(\psi^\env \land \psi_2) = (0.9607,0.9172)$.
The refinement $\psi_2$ is weaker  than $\psi_1$. Such insight into their weakness could be used to guide the refinement approach
(e.g., \cite{Alur2013,Cavezza2017})  in choosing to only refine those assumptions that may lead to weaker specifications, for instance further refining $\psi_2$ rather than  $\psi_1$.

\vspace{5pt}
\noindent
\textbf{Consistency between equally constraining formulae.} 
Consider the AMBA-AHB protocol with eight masters and the two alternative refinements: $\theta_1 = \always \eventually (\hmaster_0 \lor \lnot \hbusreq_1)$; and $\theta_2 = \always \eventually (\hmaster_1 \lor \lnot \hbusreq_2)$.
Clearly the two alternatives express the same kind of constraint on different masters. Since the two masters do not have priorities over each other, expectedly the two refinements have the same weakness:
$d(\theta^\env \land \theta_1) = d(\theta^\env \land \theta_2) = (0.9396,0.9214)$.

\vspace{2pt}
\noindent
\textbf{Performance.}
In order to compare the discriminative power of the weakness measure and implication, we perform an experiment where every pair of refinements from the refinement trees in \cite{Cavezza2017} is compared via both methods. An implication check for the pair of formulae $\phi_1$ and $\phi_2$ is performed by computing the nondeterministic transition-based generalized B\"uchi automata (TGBA) \cite{Duret-Lutz2014} of the formulae $\phi_1 \land \lnot \phi_2$ and $\phi_2 \land \lnot \phi_1$, and checking whether any of them is empty \cite{Renault2013}.

We compare the proportion of formulae pairs that have different weakness measure (and thereby can be discriminated via our proposed metric) and the proportion of formulae pairs where one formula strictly implies the other (that can be discriminated via logical implication). Table~\ref{tab:DiscriminativePower} shows the results: the columns show the total number of nodes in the refinement tree (\textbf{\#Nodes}), the corresponding number of pairs (\textbf{\#Pairs}), the percentage of pairs that can be discriminated via implication (\textbf{\%Impl}) and via weakness (\textbf{\%Weak}). The table shows that, despite weakness does not capture implication strictly in all cases, it still allows for the discrimination of a larger set of assumptions, by virtue of Axiom~\ref{axiom2}.

\begin{table}
\centering
\caption{Comparison between the discriminative power of implication and weakness}
\label{tab:DiscriminativePower}
\begin{tabular}{|c|c|c|c|c|}
	\hline 
	\textbf{Case study} & \textbf{\#Nodes ($k$)} & \textbf{\#Pairs} & \textbf{\%Impl} & \textbf{\%Weak} \\ 
	\hline 
	AMBA02 & 9 & 36 & 63.9 & 88.9 \\ 
	\hline 
	AMBA04 & 17 & 136 & 69.1 & 79.4 \\ 
	\hline 
\end{tabular} 
\end{table}

\begin{wrapfigure}[14]{r}{6.5cm}
	\vspace{-0.6cm}
	\begin{center}
		\includegraphics[width=1.1\linewidth]{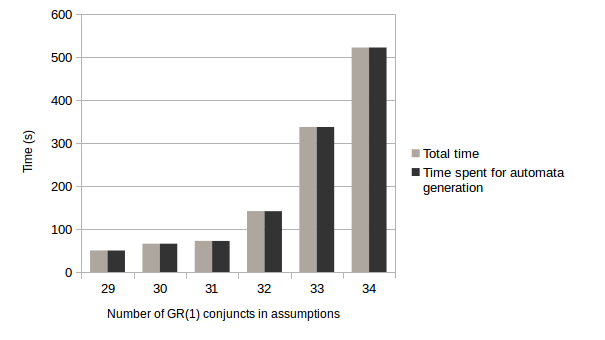}
		\caption{Execution time of weakness computation for AMBA08}
		\label{fig:resultspathtoleafamba08}
	\end{center}
\end{wrapfigure}

The time taken to compute the weakness measure for each refinement (computed via the approach in \cite{Cavezza2017}) was consistently less than 1 minute  for the lift controller, AMBA02, and AMBA04 case studies. The time needed on a representative subset of refinements from the AMBA08 example is shown in Fig.~\ref{fig:resultspathtoleafamba08} as a function of the number of GR(1) conjuncts in the assumptions.
The subset comprises a path from the root of the refinement tree (initial assumptions) to one of the 80 leaves. We observed that 79 of the 80 leaves showed similar performance as the one reported in figure; one of them, instead, took around 5200s.
Notice that over 99\% of the time is spent on DMA computation, and the remaining time is employed on eigenvalue computation.
When using implication to check whether a formula $\phi_1$ implies another formula $\phi_2$, it is necessary to produce two automata, one for $L(\phi_1) \cap L(\lnot\phi_2)$ and one for $L(\phi_2) \cap L(\lnot\phi_1)$, and then run an emptiness check on each of them. When comparing $k$ formulae, this operation must be repeated for $O(k^2)$ pairs of formulae. On the other hand, for a set of formulae containing at most $m$ fairness conditions, our weakness measure requires $m+1$ DMA computations, yielding $O(mk)$ automata needed for comparing $k$ formulae. In this respect, the advantage of our weakness measure resides in the reduced number of DMA computations with respect to implication.

The price to pay lies in the complexity of the needed automata: while weakness requires deterministic automata, implication can be checked via nondeterministic ones, which are typically faster to compute \cite{Esparza2016}. However, in the AMBA08 case we observed that the quadratic growth of implication check prevailed over the lesser complexity of nondeterministic automata: the value of $k$ for this case study is 158; while computing all weakness values for the refinement tree required a total time of 15 hours, in the same amount of time only a small fraction of the 12,403 formulae pairs could be checked for implication.

\section{Conclusion}
\label{sec:Conclusion}
In this paper we  proposed a new measure for assessing the weakness of GR(1) formulae quantitatively
and demonstrated its application in the context of weakest assumptions refinement for GR(1) controller synthesis. We  showed that strong connectedness of invariants is a sufficient requirement to guarantee that our measure distinguishes between stronger and weaker formulae in the implication sense.
We introduced a component to the measure which allows one to compare formulae with the same dimension based on the weakness of their fairness conditions. 
The major limitation of the approach is the need for deterministic automata to be produced, which induces high computation time 
because of the determinization process \cite{Esparza2016}. 

As part of our future work, we plan to explore the possibility of refining the weakness relation by including Hausdorff measure in the definition, since Hausdorff measure can distinguish between stronger and weaker \olanguages in case they are not strongly connected \cite{Merzenich:1994}. We also intend to investigate algorithms for computing---or approximating at a controlled accuracy---Hausdorff dimension on nondeterministic automata.

\subsubsection{Acknowledgments}
The support of the EPSRC HiPEDS CDT (EP/L016796/1)  is gratefully acknowledged.

\bibliographystyle{splncs03}
\bibliography{MyCollection}

\newpage

\appendix
\section*{Appendix}

\section{LTL Syntax and Semantics}
\label{app:syntax}
The syntax of LTL is defined by the following grammar:
$$\phi ::= \true \mid \false \mid p \mid \lnot \phi \mid \phi \land \phi \mid \next \phi \mid \eventually \phi \mid \always \phi \mid \phi\: \until \phi$$
where $p \in \var$.

The following statements describe LTL semantics, that is when an \oword is said to satisfy an LTL formula. Hereafter, $\phi$ and $\psi$ are LTL formulae.
\begin{align*}
w & \models \true & \text{always} \\
w & \models \false & \text{never} \\
w & \models p & \text{iff } p \in w_1 \\
w & \models \lnot \phi & \text{iff } w \not\models \phi \\
w & \models \phi \land \psi & \text{iff } w \models \phi \text{ and } w \models \psi \\
w & \models \next \phi & \text{iff } w^2 \models \phi \\
w & \models \eventually \phi & \text{iff } \exists j \in \mathbb{N} \text{ s. t. } w^j \models \phi \\
w & \models \always \phi & \text{iff } \forall j \in \mathbb{N} \: w^j \models \phi \\
w & \models \phi \until \psi & \text{iff } \exists j \in \mathbb{N} \text{ s. t. } w^j \models \psi \text{ and } \forall i<j \: w^i \models \phi
\end{align*}
In other words, $\eventually$ can be read as ``eventually'', $\always$ as ``always'', $\next$ as ``next'' and $\until$ as ``until''.

\section{Entropy, Automata and Adjacency Matrices} 
\label{app:entropy}
The entropy of a closed language can be computed on its B\"uchi automaton $\bautomaton$ if all states are accepting \cite{Merzenich:1994}. This is interpreted as a labelled graph $\mathcal{G} = (Q,E)$, where the set of \emph{nodes} $Q$ is the set of states in $\bautomaton$, $E \subseteq Q \times \mathbb{N} \times Q$ is a set of \emph{edges} such that $(q_i,n,q_j) \in E$ if and only if there exist exactly $n$ symbols $\sigma \in \Sigma$ such that $\delta(q_i,\sigma)=q_j$.

Given a subset $Q' \subseteq Q$, the \emph{subgraph} induced by $Q'$ on $\mathcal{G}$ is the graph $\mathcal{G}' = (Q',E')$ such that $(q_i,n,q_j) \in E'$ iff $(q_i,n,q_j) \in E$ and $q_i,q_j \in Q'$.

A graph is \emph{strongly connected} if for any two states $q_i,q_j \in Q$ there exists a path (a sequence of consecutive edges) from $q_i$ to $q_j$ and vice versa. If a graph $\mathcal{G}$ is not strongly connected, it can admit one or more \emph{strongly connected components} (SCCs), which are maximal strongly connected subgraphs of $\mathcal{G}$

A graph can be represented through its \emph{adjacency matrix} $A$, defined as the square matrix of size $\setsize{Q}$ whose element in position $(i,j)$ is $A_{ij} = n$ iff $(q_i,n,q_j) \in E$ and $A_{ij} = 0$ if there is no edge connecting $q_i$ and $q_j$.

The algorithm in \cite{Merzenich:1994} to compute entropy is as follows. Given a B\"uchi automaton with all states accepting and its interpretation as a graph $\mathcal{G}$
\begin{enumerate}
	\item Compute all SCCs and their adjacency matrices $A_i$;
	\item For every $A_i$ compute the maximum eigenvalue $\rho(A_i)$ (also called \emph{spectral radius} of $A_i$);
	\item Return the maximum $\rho(A_i)$.
\end{enumerate}

\section{Proof of Theorem \ref{thm:InvariantStrictWeakness}}
\label{app:ThmProof}

Invariants are special cases of safety formulae according to the classification in \cite{Manna:1992}, and therefore can be represented by B\"uchi automata where every state is accepting. First, we will construct an automaton accepting $L(\invariant_2)$ where each state keeps memory of the last symbol read. The outgoing transitions from each state are labelled only with the valuations that satisfy the invariant. Then we will show that a B\"uchi automaton of the same form can be obtained for $L(\invariant_1)$ by removing transitions and possibly states from the automaton of $L(\invariant_2)$. This yields an adjacency matrix for $L(\invariant_1)$ with a strictly lower spectral radius (maximum eigenvalue), which corresponds to Hausdorff dimension in our context.

\begin{figure}
	\centering
	\includegraphics[width=0.35\linewidth]{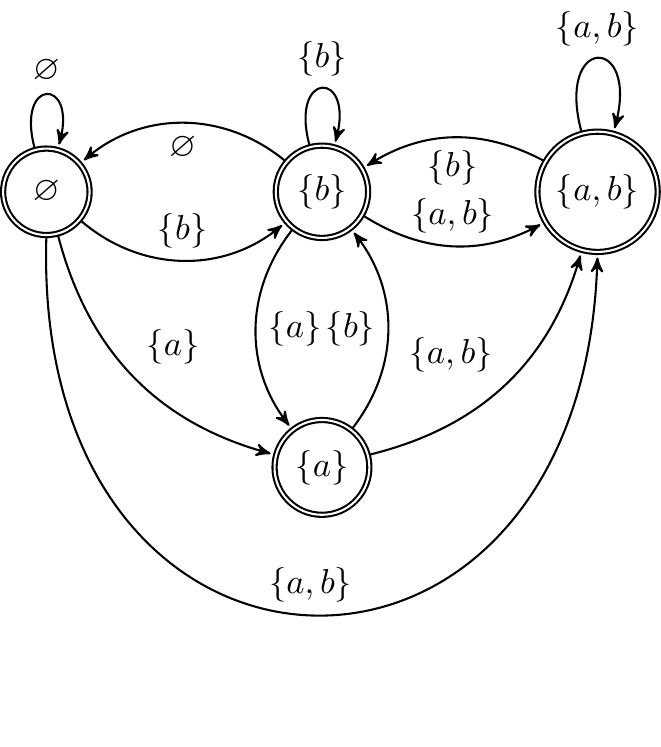}
	\caption{B\"uchi automaton of $\always(a \rightarrow \next b)$. The state labels $\lambda(q)$ are shown inside the nodes. The initial state (not shown) has transitions towards all the states in the figure, since the first symbol is unconstrained.}
	\label{fig:ExampleLabelling}
\end{figure}

As promised, first we construct the B\"uchi automata for $\invariant_1$ and $\invariant_2$.
The construction of $\bautomaton$ for $L(\invariant)$ uses a set of states $Q$ which are labelled by a one-to-one function $\lambda: Q\backslash\{q_0\} \rightarrow \Sigma$. The transition function is built such that:
\begin{enumerate}
	\item  $\delta(q,\sigma)$ is defined if and only if every \oword in $\lambda(q)\sigma \cdot \sigmaomega$ satisfies $\boolexpr{}{2}{\var \cup \next\var}$;
	\item in case $\delta(q,\sigma)$ is defined, $\lambda(\delta(q,\sigma)) = \sigma$.
\end{enumerate}
The initial state $q_0$ satisfies the following property:
\begin{enumerate}
	\item for every $\sigma \in \Sigma$, there exists $\delta(q_0,\sigma)$ if and only if there exists $\tau \in \Sigma$ such that $\sigma\tau \cdot \sigmaomega \models \boolexpression$;
	\item for every $q \in Q\backslash\{q_0\}$, if $q=\delta(q_0,\sigma)$ then $\lambda(q) = \sigma$.
\end{enumerate}
An example is pictured in Fig.~\ref{fig:ExampleLabelling}.

We show that an \oword $w$ satisfies $\invariant$ iff it corresponds to an infinite path on $\bautomaton$. Suppose $w \models \invariant$. Then for every $i \in \mathbb{N}$, $w^i \models \boolexpr{}{}{\var \cup \next \var}$. Since this formula constrains the first two symbols of $w$ only, any \oword in $w_iw_{i+1} \cdot \sigmaomega$ satisfies $\boolexpression_2$. By construction, the automaton $\bautomaton$ has a transition $\delta(q_i,w_{i+1})=q_{i+1}$ such that $\lambda(q_i)=w_i$ and $\lambda(q_{i+1})=w_{i+1}$. Therefore, if there exists a path from $q_0$ to $q_i$ induced by the prefix $w_1 \dots w_i$, there exists a path from $q_0$ to $q_{i+1}$. As a base case of the induction, consider that $w_1w_2 \cdot \sigmaomega \models \boolexpression$, and therefore there exists a path from $q_0$ to $q_1$ in $\bautomaton$.

Conversely, suppose that $w$ is an infinite sequence of transition labels such that $\bautomaton(w)=q_0q_1 \dots$. Then $\delta(q_i,w_{i+1})$ exists for every $i \in \mathbb{N} \cup \{0\}$ and $\lambda(q_i)=w_i$ for all $i \in \mathbb{N}$. By the construction of $\bautomaton$ this means that for every $i \in \mathbb{N}$ $w_iw_{i+1} \cdot \sigmaomega \models \bautomaton$, that implies $w^i \models \invariant$. Then we can conclude $w \models \invariant$. This allows us to say that $\bautomaton$ is a B\"uchi automaton of the formula $\invariant$.

If $L(\phi)$ is strongly connected, then $\bautomaton$ is also strongly connected, except for the initial state. Let $q_n = \delta(q_0,w) = \delta(\delta(\delta(\dots \delta(q_0,w_1), \dots), w_{n-1}), w_n)$ be the state reached by $\bautomaton$ after reading the prefix $w \in \prefnlanguage{n}{L(\phi)}$, and $q_m = \delta(q_0,v)$ for $v \in \prefnlanguage{m}{L(\phi)}$. By construction, $\lambda(q_n)=w_n$ and $\lambda(q_m)=v_m$. Since $L(\phi)$ is strongly connected, there exist $v' \in \sigmastar$ such that $S_{vv'} = L(\phi)$. Therefore $vv'w \in \preflanguage{L(\phi)}$ and $\lambda(q_0,vv'w) = w_n$, so $\delta(q_0,vv'w) = q_n$. So, there exists a path in $\bautomaton$ from $q_m$ to $q_n$. Symmetrically, there exists a path from $q_n$ to $q_m$. We have proved that for any pair of states reachable from $q_0$ there is a path between them in both directions, that is the graph induced by non-initial states is strongly connected.

Now consider the automata $\bautomaton_1$ and $\bautomaton_2$ of $L(\invariant_1)$ and $L(\invariant_2)$ respectively. Since the two formulae are pure invariants, $\invariant_1 \rightarrow \invariant_2$ if and only if $\boolexpr{}{1}{\var \cup \next \var} \rightarrow \boolexpr{}{2}{\var \cup \next \var}$. So, given the hypothesis that $\invariant_1$ is strictly stronger than $\invariant_2$ there must exist a pair of valuations $\sigma,\tau$ such that $\sigma\tau \cdot \sigmaomega \models \invariant_2$ but $\sigma\tau \cdot \sigmaomega \not\models \invariant_1$. By construction this corresponds to at least one transition $\delta(q,\tau)$ that exists in $\bautomaton_2$ and does not in $\bautomaton_1$. Consequently, we can conclude that $\bautomaton_1$ is a proper subgraph of $\bautomaton_2$.

The next step is to construct the adjacency matrices corresponding to $\bautomaton_1$ and $\bautomaton_2$ excluding the respective initial states. Let $Q_1\backslash\{q_{0,1}\}$ and $Q_2\backslash\{q_{0,2}\}$ be the set of non-initial states of $\bautomaton_1$ and $\bautomaton_2$, respectively, and $\delta_1$ and $\delta_2$ their respective transition functions. Let $A$ and $B$ be the adjacency matrices of the graphs $(Q_1\backslash\{q_0\},\delta_1)$ and $(Q_2\backslash\{q_0\},\delta_2)$. Consider that by construction all transitions between two states are labelled by exactly one valuation: so, each element of these two matrices is either a 0 or a 1.

Since the transitions of $\delta_2$ are a proper subset of the transitions of $\delta_1$, we have for each element $(i,j)$ $A_{ij} \leq B_{ij}$, and $A_{kh} < B_{kh}$ for some $(k,h)$, that is to say $A_{kh}=0, B_{kh}=1$. Since $\invariant_2$ is strongly connected, its adjacency matrix $B$ is irreducible (for more details on the correspondence between strongly connected digraphs and irreducible matrices see Chapter 6 of \cite{Horn:1985}). Moreover, $A+B$ is also irreducible, since it is the sum of two nonnegative matrices one of which is irreducible. We can therefore apply the property stated in Chapter 2, Corollary 1.5 of \cite{Berman:1994}, which guarantees that under the given conditions $\rho(A) < \rho(B)$.

Taking the logarithm on both sides, we get $\Hausdim{\invariant_1} < \Hausdim{\invariant_2}$, finishing the proof. \qed

\section{Proof of Theorem~\ref{thm:HausdimComputation}} 
\label{app:thm_hausdimcomp}
The language of $\phi^c = \always \boolexpr{\inv}{}{\var \cup \next\var} \land \eventually \always \lnot \boolexpr{\fair}{}{\var}$ is not closed. We will therefore build a Muller automaton $\mautomaton$ for this language and show that applying the algorithm of Sec.~\ref{sec:WeaknessHausDim} is equivalent to computing the Hausdorff dimension of the \olanguage $L(\always (\boolexpr{\inv}{}{\var \cup \next\var}\land \lnot \boolexpr{\fair}{}{\var}))$. We are supposing that $\boolexpr{\inv}{}{\var \cup \next\var}$ and $\lnot \boolexpr{\fair}{}{\var}$ are consistent.

Let us first construct a B\"uchi automaton for $\invariant = \always \boolexpr{\inv}{}{\var}$ as in Appendix~\ref{app:ThmProof}. We have shown that any infinite path on this automaton satisfies $\invariant$. We now replace the B\"uchi winning condition $F$ with a Muller condition $T$ that accounts for satisfying $\cfairness = \eventually \always \lnot \boolexpr{\fair}{}{\var}$. Let us denote by $Q_{\cfair}$ the set of states $q$ such that $\lambda(q)$ satisfies $\lnot \boolexpr{\fair}{}{\var}$. The accepting table $T$ is then $T := 2^{Q_{\cfair}}$.

It is clear that an \oword $w$ satisfies $\phi^c$ if and only if $w$ is accepted by $\mautomaton$. Suppose $w$ is accepted by $\mautomaton$. Then for some $S' \in T$, $\Inff(w) = S'$. Since for each $q \in S'$ $\lambda(q)$ satisfies $\lnot \boolexpr{\fair}{}{\var}$, by construction the valuations $w_i$ leading to $q$ satisfy $\lnot \boolexpr{\fair}{}{\var}$. Therefore we conclude that there exists an infinite suffix of $w$ that satisfies $\always \lnot \boolexpr{\fair}{}{\var}$, that is $w \models \cfairness$. Moreover, $w$ induces an infinite path on the automaton $\mautomaton$, and therefore by construction $w \models \invariant$. Therefore, $w \models \phi^c$.

Conversely, suppose $w$ satisfies $\phi^c$. Then it satisfies $\invariant$, and thereby induces an infinite path over $\mautomaton$. Moreover, it satisfies $\cfairness$, and therefore there exists a suffix of $w$ that satisfies $\always \lnot \boolexpr{\fair}{}{\var}$. So, by construction $\Inff(w) \subseteq Q_{\cfair}$, that is $\Inff(w) \in T$.

The algorithm in Sec.~\ref{sec:WeaknessHausDim} requires to compute the Hausdorff dimension of every language $C_{S'}$ for $S' \in T$. The Hausdorff dimension of $L(\phi^c)$ is the maximum of such Hausdorff dimensions. The B\"uchi automaton of the (closed) language $C_{S'}$ corresponds to the subgraph induced by the states in $S'$ onto $\mautomaton$, with any state being accepting \cite{Staiger:1998}. The maximum Hausdorff dimension is attained for $S' = Q_{\cfair}$.

By construction, $\lambda(q)$ satisfies $\lnot \boolexpr{\fair}{}{\var}$ for every $q \in Q_{\cfair}$, and for every pair of consecutive states $(q,q')$ we have $\lambda(q)\lambda(q')\cdot\sigmaomega \models \boolexpr{\inv}{}{\var\cup\next\var}$. Therefore, the subgraph induced by $Q_{\cfair}$ corresponds to the B\"uchi automaton of the closed language $L(\always(\boolexpr{\inv}{}{\var\cup\next\var} \land \lnot \boolexpr{\fair}{}{\var}))$.

In conclusion,
$$d_2(\phi)=\Hausdim{L(\phi^c)}=L(\always(\boolexpr{\inv}{}{\var\cup\next\var} \land \lnot \boolexpr{\fair}{}{\var})),$$
finishing the proof. \qed

\section{Quantitative Model Checking: a Further Application Example}
\label{app:Evaluation}
In this section we provide an application of computing Hausdorff dimension on fairness complements in a quantitative model checking example from \cite{Asarin:2014}.

Applied to the model checking problem, our weakness measure extends the quantitative approach in \cite{Asarin:2014} to fairness properties. Consider the Dining Philosophers problem with three philosophers. Let $\phi^{\DP}$ be the GR(1) formula describing all the philosophers' behaviors that do not reach the deadlock state. The goal is to assign a measure to the subset of these behaviors such that none of the philosophers starve. This condition is expressed by the fairness formula $\phi^{\DP}_{\fair} = \always \eventually (\stateone=\EAT) \land \always \eventually (\statetwo=\EAT) \land \always \eventually (\statethree=\EAT)$. The assigned measure is meant to characterize the degree of satisfaction of this formula by a model of the Dining Philosophers problem. When using entropy only, $\entropy{L(\phi^{\DP})} = \entropy{L(\phi^{\DP} \land \phi^{\DP}_{\fair})} = 0.0718$. When using our two-component measure function, the degree of satisfaction of a fairness formula is measured indirectly through the subset of behaviors excluded by the formula itself. The result is $d(\phi^{\DP}) = (0.0718,0)$, $d(\phi^{\DP} \land \phi^{\DP}_{\fair}) = (0.0718,0.0479)$. Therefore, contrary to the work in \cite{Asarin:2014}, our measure is able to capture the difference in the restrictiveness of the two formulae.

\section{Specification of the Extended Lift Example}
\label{app:ExtLift}
Assumptions:
\begin{enumerate}
	\item $\lnot b_1 \land \lnot b_2 \land \lnot b_3 \land \lnot \alarm$
	\item $\always((b_1 \land f_1) \rightarrow \next \lnot b_1)$
	\item $\always((b_2 \land f_2) \rightarrow \next \lnot b_2)$
	\item $\always((b_3 \land f_3) \rightarrow \next \lnot b_3)$
	\item $\always((b_1 \land \lnot f_1) \rightarrow \next b_1)$
	\item $\always((b_2 \land \lnot f_2) \rightarrow \next b_2)$
	\item $\always((b_3 \land \lnot f_3) \rightarrow \next b_3)$
\end{enumerate}
Guarantees:
\begin{enumerate}
	\item $f_1 \land \lnot f_2 \land \neg f_3 \land \lnot \stopp$
	\item $\always(\lnot (f_1 \land f_2) \land \lnot(f_2 \land f_3) \land \lnot (f_1 \land f_3))$
	\item $\always((\lnot \stopp \land f_3) \rightarrow \next (f_2 \lor f_3))$
	\item $\always((\lnot \stopp \land f_1) \rightarrow \next (f_1 \lor f_2))$
	\item $\always(((f_1 \land \next f_2) \lor (f_2 \land \next f_3) \lor (f_3 \land \next f_2) \lor (f_2 \land \next f_1)) \rightarrow (b_1 \lor b_2 \lor b_3))$
	\item $\always\eventually ((\lnot \stopp \land b_1) \rightarrow f_1)$
	\item $\always\eventually ((\lnot \stopp \land b_2) \rightarrow f_2)$
	\item $\always\eventually ((\lnot \stopp \land b_3) \rightarrow f_3)$
	\item $\always(\stopp \rightarrow ((f_1 \rightarrow \next f_1) \land (f_2 \rightarrow \next f_2) \land (f_3 \rightarrow \next f_3)))$
	\item $\always(\alarm \rightarrow \next \stopp)$
	\item $\always\eventually f_1$
	\item $\always\eventually f_2$
	\item $\always\eventually f_3$
\end{enumerate}

\end{document}